\documentclass[preprint,showpacs,preprintnumbers,amsmath,amssymb]{revtex4}
\usepackage{epsfig,amsmath,amssymb,graphics,color,calc,wasysym}

\newcommand{\be}{\begin{equation}}
\newcommand{\ee}{\end{equation}}
\newcommand{\ba}{\begin{eqnarray}}
\newcommand{\ea}{\end{eqnarray}}

\renewcommand{\phi}{\varphi}

\begin{document}

\title{Collective Diffusion of Colloidal Hard Rods in Smectic Liquid Crystals: Effect of Particle Anisotropy}

\author{Alessandro Patti\footnote{a.patti@uu.nl}}
\affiliation{Soft Condensed Matter Group, Debye Institute for NanoMaterials Science, Utrecht University, Princetonplein 5, 3584 CC, Utrecht, The Netherlands}

\author{Djamel El Masri}
\affiliation{Soft Condensed Matter Group, Debye Institute for NanoMaterials Science, Utrecht University, Princetonplein 5, 3584 CC, Utrecht, The Netherlands}

\author{Ren\'{e} van Roij}
\affiliation{Institute for Theoretical Physics, Utrecht University, Leuvenlaan 4, 3584 CE, Utrecht, The Netherlands}

\author{Marjolein Dijkstra\footnote{m.dijkstra1@uu.nl}}
\affiliation{Soft Condensed Matter Group, Debye Institute for NanoMaterials Science, Utrecht University, Princetonplein 5, 3584 CC, Utrecht, The Netherlands}

\date{\today}

\begin{abstract}
We study the layer-to-layer diffusion in smectic-A liquid crystals of colloidal hard rods  with  different length-to-diameter ratios using computer simulations. The layered arrangement of the smectic phase yields a hopping-type diffusion due to the presence of permanent barriers and transient cages.  Remarkably, we detect stringlike clusters composed of inter-layer rods moving cooperatively along the nematic director.  Furthermore, we find that the structural relaxation in \textit{equilibrium} smectic phases shows interesting similarities with that of \textit{out-of-equilibrium} supercooled liquids, although there the particles are kinetically trapped in transient rather than permanent cages. Additionally, at fixed packing fraction we find that the barrier height increases with increasing particle anisotropy, and hence the dynamics is more heterogeneous and non-Gaussian for longer rods, yielding a lower diffusion coefficient along the nematic director and smaller clusters of inter-layer particles that move less cooperatively. At fixed barrier height,  the dynamics becomes more non-Gaussian and heterogeneous for longer rods that move more collectively giving rise to  a higher  diffusion coefficient along the nematic director.
\end{abstract}

\pacs{82.70.Dd; 61.30.-v; 87.15.Vv}

\maketitle

\section{INTRODUCTION}
\label{sec:INTRODUCTION}

Liquid crystals (LCs) are states of matter whose properties are in between those  of a crystalline solid and  an isotropic liquid phase \cite{chandrasekhar}. They are usually classified in terms of positional and orientational order. Nematic LCs exhibit long-range orientational order, as the anisotropic particles are on average aligned along a preferred direction, but they lack long-range positional order. Smectic phases consist of stacks of fluid-like layers of orientationally ordered particles, where each layer is often considered to be a two-dimensional fluid. Onsager showed in his seminal contribution  the existence of a purely entropy-driven isotropic-to-nematic (I-N) transition in a system of infinitely long hard rods \cite{onsager}. Moreover, the I-N transition was confirmed by computer simulations for systems of hard rods with {\em finite} length. Additionally, Frenkel and coworkers explored the formation of smectic LCs of perfectly aligned \cite{stroobants} and freely rotating \cite{frenkel} hard rods, and found a thermodynamically stable smectic phase of hard rods as a result of entropic  effects. The equilibrium properties of smectic LCs are well-studied and are well-understood by now \cite{degennes}. Experimental \cite{dogic1, dogic2, dogic3}, theoretical \cite{mulder, wen, somoza, holyst, vanderschoot, bolhuis1}, and computational \cite{veerman, polson} studies have analyzed the phase behavior and structure of smectic LCs of  colloidal hard rods. Other investigations involve extensions to binary mixtures with rods of different geometry \cite{vroege,cinacchi1,martinezraton,roij1996,roij1996b,roij1998,dijkstra1997}, with other anisotropic \cite{galindo,roij1994,camp1996,camp1997} or spherical  colloidal particles \cite{adams1, adams2, vliegenhart, dogic4}, or with non-adsorbing polymer as depletants \cite{dogic5,bolhuis2,savenko}.

By contrast, the dynamics on a single-particle level in smectic LCs have only recently received  attention, although an early study on the diffusion (or "permeation") of anisotropic particles had already been reported more than forty years ago by Helfrich in order
to explain the capillary flow in cholesteric and smectic LCs \cite{helfrich}. Substantial advances in new experimental techniques (e.g. NMR coupled to strong magnetic field gradients \cite{furo} or fluorescent labeling \cite{lettinga}) disclosed the non-Gaussian
nature and quasi-quantized behavior of the layer-to-layer diffusion. These achievements sparked off new theoretical work, based on dynamic density functional theory, which not only confirmed the non-Gaussian layer-to-layer hopping-type diffusion and the presence of permanent barriers due to the static smectic background, but also showed the relevance of temporary cages due to the mutual trapping of neighboring particles \cite{bier, grelet}.  We note that non-Gaussian dynamics due to a rattling-and-jumping diffusive behavior is common in two-dimensional liquids \cite{hurley}, cluster crystals \cite{moreno}, and glasses \cite{chaudhuri}, and has also been observed for the diffusion of a single particle in a periodic external potential \cite{vorselaars}. It is therefore not surprising to observe similar behavior in smectic LCs. Our simulations of parallel \cite{matena} and freely rotating \cite{patti} hard rods supported indeed these conclusions, but  unveiled in addition a striking  analogy with the non-exponential structural
relaxation and non-Gaussian dynamics of supercooled liquids. The non-exponential relaxation of the density fluctuations might be due to either a \textit{heterogeneous scenario} with particles relaxing exponentially at different relaxation rates, or a
\textit{homogeneous scenario} with particles relaxing non-exponentially at very similar rates \cite{richert}. Here, we investigate  the collective motion of fast-moving rods in stringlike clusters. Cooperative diffusion, which accounts for the non-exponential
decay of the correlation functions, yields an intriguing link between the dynamics observed in \textit{equilibrium} smectic LCs and that of \textit{out-of-equilibrium} supercooled liquids. As far as smectic LCs are concerned, this remarkable collective motion could not be captured by the one-particle analysis of Ref. \cite{vorselaars}, and was not observed in Ref. \cite{cinacchi}. Two-dimensional liquids of soft disks \cite{hurley} and cluster crystals \cite{moreno} did not show this feature either. By contrast, several experimental and computational studies on glassy systems reported the existence of structural heterogeneities \cite{donati, marcus, giovambattista, aichele, teboul}. In particular, Glotzer and coworkers performed molecular dynamics simulations on a \textit{fragile} glass-forming liquid and detected cooperative motion of stringlike clusters with an increasing string length of up to $\sim 15$ particles by cooling the system towards the glass transition \cite{donati}. Similar results were observed more recently in silica, a \textit{strong} glass-former, suggesting that stringlike motion is a universal property of supercooled liquids \cite{teboul}.

In this paper, we investigate  the effect of anisotropy of the rods on the non-Gaussian layer-to-layer diffusion and cooperative motion of stringlike clusters in bulk smectic LCs of freely rotating hard rods.  In supercooled liquids, it is generally accepted that in the case of  heterogeneous dynamics, the cooperative motion of particle clusters plays a crucial role in the structural relaxation  \cite{darst}. We argue that a similar behavior can be observed in smectic LCs, where cooperative layer-to-layer motion of strings with various sizes contributes to the long-time relaxation behavior of the smectic phase.

This paper is organized as follows. In section II, we introduce the model, the simulation details, and the computational tools to describe the layer-to-layer diffusion. The results on the non-Gaussian and heterogeneous dynamics, as well as  evidence of cooperative motion are discussed in section III. In the last section, we present our conclusions.

\section{MODEL AND SIMULATION METHODOLOGY}
\label{sec:MODELANDSIMULATIONMETHODOLOGY}

We perform simulations of a system with $N=1530-3000$ freely rotating hard spherocylindrical rods with diameter $D$ and length $L+D$, distributed over 5-10 smectic layers of approximately 300 rods each. Three different aspect ratios, $L^{*}=L/D$, are considered: $L^{*}=3.4, 3.8,$  and $5.0$. The region of stability of the smectic phase decreases with $L^{*}$, and disappears at $L^{*} \leq 3.1$, where only a stable isotropic-crystal phase transition is found \cite{bolhuis1}. For $L^{*}=3.8$ and $5.0$, the smectic phase is stable for $2.3 \leq P^{*} \leq 2.8$ and  $1.4 \leq P^{*} \leq 2.3$, respectively, where $P^{*}= \beta PD^{3}$ is the reduced pressure. $\beta$ is $1/k_{B}T$, with $k_{B}$ the Boltzmann's constant and $T$ the temperature. For lower $P^*$, the smectic phase transforms into a nematic phase, while for higher $P^*$ the smectic freezes into a crystal phase. For $L^{*}=3.4$, the nematic phase is unstable, and the smectic phase transforms directly into an isotropic phase for $P^{*} \leq 2.8$ and crystallizes for $P^{*} \geq 3.0$. We studied the dynamics of the bulk smectic phase at the pressures and packing fractions indicated in Table I. For convenience, we label the systems with $S_{1}-S_{6}$ (see Table I).

\begin{table}[!h]
\centering
\begin{tabular} {|p{0.7in}|c|c|c|c|c|c|} \hline
$L^{*}$ 	& 	\multicolumn{2}{|c|}{3.4}						&	 \multicolumn{2}{|c|}{3.8}&		\multicolumn{2}{|c|}{5.0}\\ \hline
Stable Sm	&	\multicolumn{2}{|c|}{$2.8 \leq P^{*} \leq 3.0$}	&	 \multicolumn{2}{|c|}{$2.3 \leq P^{*} \leq 2.8$}	&	 \multicolumn{2}{|c|}{$1.4 \leq P^{*} \leq 2.3$}	\\\hline
$P^{*}$ 					&2.85 		&3.00 		&2.35 		&2.50 		&1.60 		&2.00 	\\ \hline
$\eta$					&0.556		&0.568		&0.536		&0.551		&0.508		&0.557	\\ \hline
$h/(L+D)$ 				&1.018		&1.014		&1.023		&1.015		&1.048		&1.030	\\ \hline
$\sigma/(L+D)$			&0.050		&0.043		&0.075		&0.052 		&0.093		&0.043	\\ \hline
$D^L_{z}\tau/D^{2}$  		&0.036		&0.020		&0.116		&0.050		& 0.325		&0.028	\\ \hline
$D^L_{xy}\tau/D^{2}$		&0.230		&0.179		&0.295		&0.256		&0.413		&0.268	\\ \hline
$\textit{\={t}}/\tau$			&0.19		&0.23		&0.17		&0.20		&0.14		&0.20	\\ \hline
$t_{J}^{*}/\tau$			&0.28		&0.30		&0.27		&0.32		&0.20		&0.27	\\ \hline
$t_{J}^{max}/\tau$			&1.10		&1.35		&1.40		&1.65		&1.00		&1.20	\\ \hline
$\textit{\={f}}_{c}$		&0.04		&0.02		&0.04		&0.04		&0.09		&0.03	\\ \hline
$f^{*}_{c}$ 				&0.04		&0.02		&0.05		&0.05		&0.09		&0.03	\\ \hline
$f^{max}_{c}$			&0.06		&0.04		&0.14		&0.07		&0.25		&0.04	\\ \hline
$U_{0}/k_{B}T$			&6.2		&7.5		&4.0		&5.8		&3.5		&7.5	\\ \hline
						&$S_{1}$ 	&$S_{2}$	&$S_{3}$	&$S_{4}$	&$S_{5}$	&$S_{6}$ \\\hline
\end{tabular}
\caption{Details of the systems that we studied in this paper, consisting of hard spherocylinders with varying length-to-diameter ratio $L^*=L/D$ and reduced pressures $P^{*}=\beta P D^3$, and corresponding packing fractions $\eta$. For comparison, we  give the pressure range of the stable smectic phase for the corresponding systems. Additionally, we give the layer spacing (\textit{h}); the standard deviation of the displacement from the equilibrium smectic phase ($\sigma$) in units of \textit{(L+D)}; the long-time in-layer $D^L_{xy}$ and inter-layer $D^L_{z}$ diffusion coefficients in units of $\tau/D^{2}$; the most probable ($\textit{\={t}}$), median ($t_{J}^{*}$), and maximal ($t_{J}^{max}$) jump times; the fraction of collective jumps $\textit{\={f}}_{c}$, $f^{*}_{c}$, and $f^{max}_{c}$, calculated with a temporal interval $\Delta t_0 $ equal to $\textit{\={t}}$, $ t_J^*$, and $t_J^{max}$, respectively; and finally, the height of the energy barriers ($U_{0}$) in units of $k_{B}T$. The systems are labeled by $S_{1}-S_{6}$.}
\end{table}

We performed Monte Carlo (MC) simulations in a rectangular box of volume \textit{V} with periodic boundary conditions. Firstly, we performed equilibration runs in the isobaric-isothermal (\textit{NPT}) ensemble to expand the system from an ordered crystalline phase to an equilibrated smectic phase. Each MC cycle consisted of \textit{N} attempts to displace and/or rotate the randomly selected particles, plus an attempt to change the box volume by modifying the three box lengths independently. Translational and rotational moves were accepted if no overlap was detected. The systems were considered to be equilibrated when the packing fraction reached a  constant value within the statistical fluctuations. A typical equilibration run took roughly $3 \times 10^{6}$ MC cycles and was followed by a production run in the isochoric-isothermal (\textit{NVT}) ensemble to analyze the relaxation dynamics. At this stage, we kept the volume constant to avoid unphysical collective moves which do not mimic the Brownian dynamics of the rods properly. Standard MC simulations with small displacements were used to mimic Brownian motion. This computational approach was shown to be very efficient to study the slow relaxation of glasses at low temperatures \cite{berthier} or at high concentrations \cite{pfleiderer}. We fixed the maximum displacement according to (\textit{i}) a reasonable time of simulation, (\textit{ii}) a satisfactory acceptance rate, and (\textit{iii}) a suitable description of the Brownian motion of colloidal particles suspended in a fluid (see Fig. 1). To this end, we monitored the mean-square displacement in the \textit{z} and \textit{xy} directions for several values of the maximum step size $\delta_{max}$, with $\delta_{max, z}=2\delta_{max, xy}$ to take into account the anisotropy of the self-diffusion of the rods \cite{doi}. We found $\delta_{max, xy}=D/10$ and $\delta_{max, z}=D/5$ to be the optimal values which satisfied the above requirements.

\begin{figure}[!ht]
\center
\includegraphics[width=0.48\textwidth]{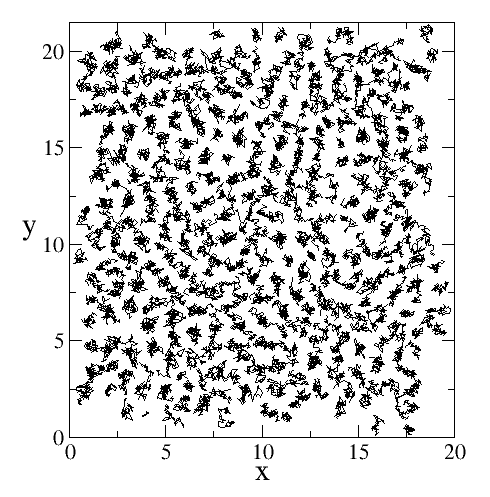}
\caption{Trajectories in a plane perpendicular to the nematic director of approximately 300 rods in a smectic layer of hard spherocylinders with a length-to-diameter ratio $L^*=5.0$ and reduced pressure $P^*=1.60$ collected over $5\times 10^{3}$ MC cycles. }
\end{figure}

As unit of time, we have chosen $\tau\equiv D^{2}/D^s_{tr}$, where $D^s_{tr}$ is the short-time translational diffusion coefficient at $L^{*}=5.0$, which is the isotropic average of the diffusion coefficients in the three space dimensions: $D^s_{tr} \equiv (D^s_{z}+2D^s_{xy})/3$. At short times, when the single particle is rattling around its original position without feeling the presence of its surrounding neighbors, the dependence of $D^s_{tr}$ on the pressure can be safely neglected. We checked that our results (measured in units of $\tau$) were independent of $\delta_{max}$.

In order to characterize the layer-to-layer hopping-type diffusion and the structural relaxation of our systems, we calculated (\textit{i}) the energy barrier, (\textit{ii}) the self part of the van Hove correlation function, (\textit{iii}) the non-Gaussian parameter, (\textit{iv}) the mean square displacement, (\textit{v}) the intermediate scattering function, (\textit{vi}) the probability distribution of the size of the stringlike clusters, and (\textit{vii}) their dynamic cooperativity.

\textit{Energy barrier}. We computed the energy barriers from the (relative) probability $\pi(z)$ of finding a rod at a given position \textit{z} along the nematic director $\hat{n}$. As reported in Ref. \cite{lettinga}, this probability is proportional to the Boltzmann factor

\begin{equation}
 \pi(z) \propto \exp \left[ -U(z)/k_{B}T \right],
\end{equation}

\noindent where $U(z)$ denotes the effective potential for the layer-to-layer diffusion.

\textit{Self-part of the van Hove correlation function.} To quantify the heterogeneous dynamics due to the rattling and hopping type $z$-diffusion of the rods, we calculate the  self-part of the van Hove correlation function (VHF), which  measures the probability distribution for the displacements of the rods along the nematic director $\hat{n}$ at time $t_{0}+t$, given their \textit{z} positions at $t_{0}$. It is defined as \cite{hansen}

\begin{equation}
G_{s}(z, t)=\frac{1}{N}\left\langle \sum_{i=1}^N \delta \left[ z - \left( z_{i}(t_{0}+t)-z_{i}(t_{0}) \right) \right] \right\rangle
\end{equation}

\noindent with $\left\langle ... \right\rangle$ the ensemble average over all particles and initial time $t_{0}$, and $\delta$ is the Dirac-delta. Note that $G_{s}(z, t)$ would be a Gaussian distribution of \textit{z} for freely diffusive particles.

\textit{Non-Gaussian parameter.} A quantitative description of the non-Gaussian behavior of the layer-to-layer diffusion can be obtained in terms of the non-Gaussian parameter (NGP) \cite{rahman}:

\begin{equation}
\alpha_{2, z}(t)=\frac{\left\langle \Delta z(t)^{4}\right\rangle}{(1+2/d)\left\langle \Delta z(t)^{2} \right\rangle^{2}}-1,
\end{equation}

\noindent where $\Delta z(t)=z(t_{0}+t)-z(t_{0})$ denotes the \textit{z}-displacement of a rod in the time interval $t$ starting at $t_{0}$, and $d=1$. Heterogeneous dynamics occurs on a time scale \textit{t}, if the NGP is non-vanishing. For the in-layer diffusion, a similar NGP, $\alpha_{2, xy}(t)$, with $d=2$ can be defined.

\textit{Intermediate scattering function.} The structural relaxation and, in particular, the decay of the density fluctuations can be quantified by the self-part of the intermediate scattering function (ISF)

\begin{equation}
F_{s}(t)=\left\langle \exp\left[ i\textbf{q} \cdot \Delta \textbf{r}(t)\right]  \right\rangle
\end{equation}

\noindent at the wave vectors $\textbf{q}=(0, 0, q_{z})$ and $(q_{x}, q_{y}, 0)$ corresponding to the main peaks of the static structure factor in the \textit{z} and \textit{xy} directions, respectively.

\textit{Cluster size distribution}. In order to investigate whether or not there is cooperative motion of stringlike clusters of particles, which might be responsible for the heterogeneous dynamics and stretched-exponential decay of the ISF, we first determine the clusters of fast-moving particles. The fast-moving particles can be identified as those particles that have traveled  a substantially longer distance than the average in a certain time interval and are intimately related to the particles that reside in between the center-of-masses of two smectic layers. Hence, particles  that  reside more than some \textit{rattling} distance $\delta_{rat}$ from the center-of-mass of the nearest smectic layer in a static configuration are defined as interlayer rods. To  define $\delta_{rat}$, we first calculate the variance of one period of $\pi(z)$, defined as

\begin{equation}
\sigma^{2} \equiv \int_{-h/2}^{h/2} z^{2} \pi(z) dz,
\end{equation}

\noindent where \textit{h} is the layer spacing obtained from fits to the density profiles $\pi(z)$, and $\int_{-h/2}^{h/2} \pi(z) dz=1$. We assume that two interlayer rods belong to the same string if their \textit{z} and \textit{xy} distances are smaller than \textit{h} and \textit{D}, respectively. We calculated the probability distribution $P(n)$ of the number of rods  \textit{n} in a string for $\delta_{rat}=2\sigma$. In Table I, we give the values of the layer spacing \textit{h} and the standard deviation (square root of the variance) $\sigma$ for the systems that we studied. We find that denser states reveal a slightly smaller layer spacing  for all aspect ratios, as expected.

\textit{Dynamic cooperativity.} To check whether or not the rods in a given string move collectively from one layer to another layer, we require a cluster criterion involving a spatial proximity of particles and a temporal proximity of jumping rods. To this end, we assume that two jumping rods \textit{i} and \textit{j} are moving cooperatively if (1) their arrival times $t^{(i)}$ and $t^{(j)}$ in their new layers (i.e. the time at which their distance to the middle of the new layer equals $\delta_{rat}$) satisfy $\vert t^{(i)} - t^{(j)} \vert < \Delta t_{0}$, and (2) the center-of-mass positions of the two particles, ${\bf r}_i(t^{(i)})$ and  ${\bf r}_j(t^{(j)})$,  satisfy the above mentioned static cluster criterion. The time interval $\Delta t_{0}$ is fixed on the basis of the distribution of jump times and will be introduced in the following section.

\section{RESULTS}
\label{sec:RESULTS}

The layered structure of smectic LCs yields an effective periodic potential $U(z)$ for  the diffusion of  rods out of the middle of a smectic layer to another layer. The permanent energy barriers for the layer-to-layer diffusion are determined from the effective potential $U(z)$ introduced in Eq. (1). In order to evaluate the effect of the pressure on the inter-layer diffusion, two separate state points are considered for each aspect ratio: one in the proximity of the nematic-smectic (N-Sm) or isotropic-smectic (I-Sm) transition, and the other in the proximity of the smectic-crystal (Sm-K) transition. In Fig. 2, we show $U(z)$ for the six systems along with the fit $U(z)=\sum_{i=1}^{m} U_{i}\left[ \sin(\pi z/h)\right] ^{2i}$, where \textit{m} is an integer number, $U_{0}\equiv \sum_{i=1}^{m}U_{i}$ is the barrier height, and $h$ the interlayer spacing given in Table I. 

\begin{figure}[!ht]
\center
\includegraphics[width=0.48\textwidth]{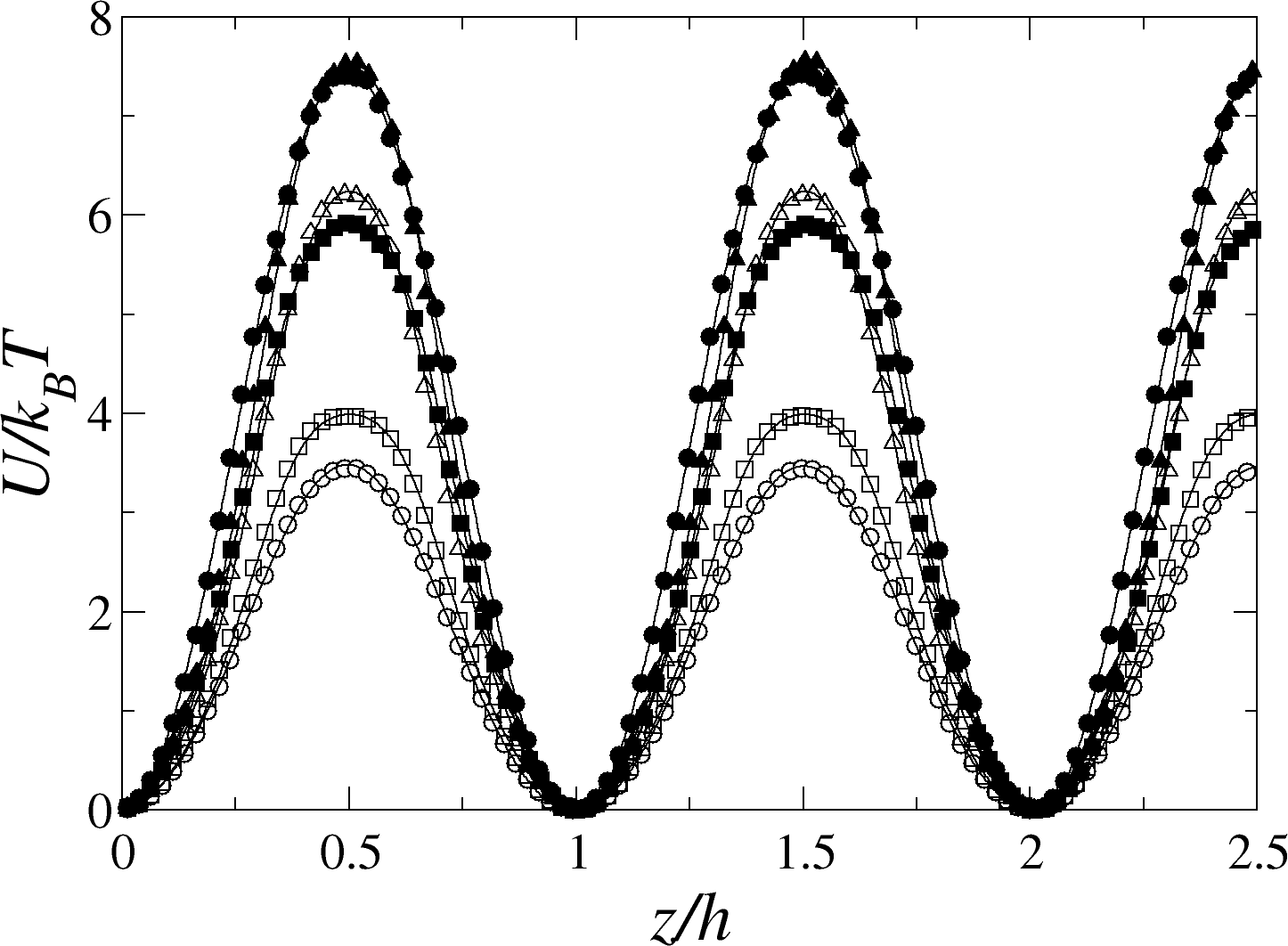}
\caption{Potential energy barriers for the layer-to-layer diffusion in smectic LCs of hard spherocylinders with varying length-to-diameter ratios $L^*$ and reduced pressures $P^*$. The open symbols refer to weak smectic states $L^*=3.4;P^*=2.85$  ($\vartriangle$), $L^*=3.8;P^*=2.35$ ($\square$), and $L^*=5.0;P^*=1.60$ (\Circle). The solid symbols denote state points deep in the smectic phase $L^*=3.4;P^*=3.00$ ($\blacktriangle$),  $L^*=3.8;P^*=2.50$ ($\blacksquare$), and $L^*=5.0;P^*=2.00$  (\CIRCLE). The solid lines are fits with $m=5$ harmonic modes.}
\end{figure}

As a general consideration, we observe that  the energy barriers increase with increasing packing fraction. The denser state is  characterized by a much stronger confinement of the particles to the middle of the smectic layers. This is especially evident at $L^{*}=5.0$ as the height of the barrier increases from $3.5k_{B}T$ to $7.5k_{B}T$ from $\eta=0.508$ to 0.557. A similar behavior was detected in experiments on smectic LCs of \textit{fd} viruses, where the height of the energy barriers was found to increase from $0.66k_{B}T$ to  $1.36k_{B}T$  by decreasing the ionic strength \cite{lettinga}. According to the  authors, a low ionic strength gives rise to stronger correlations due to more pronounced electrostatic interactions between the virus particles, resulting in higher energy barriers. Older experiments on thermotropic liquid crystals estimated energy barriers of $\sim$1-4$k_{B}T$ \cite{volino, richardson}. Our results are in  good quantitative agreement with those obtained by computer simulations in Ref. \cite{vanduijneveldt}, where rods with  $L^{*}=3.8$ and 5.0 were studied. Additionally,  these authors found  that the energy barrier for a rod to achieve a transverse inter-layer position was several $k_{B}T$ higher than that to diffuse from layer to layer, confirming the difficulty to detect transverse particles in between smectic layers \cite{vanroij}. We further observe that the packing fraction is not the only parameter affecting the effective potentials $U(z)$ as displayed in Fig. 2. It is interesting to note that the two systems $S_{1}$ and $S_{6}$, with $L^*=3.4$ and $ P^*=2.85$ as denoted by the open triangles ($\vartriangle$) and with $L^*=5.0$ and $P^*=2.00$ as shown by the solid circles (\CIRCLE), corresponding to very similar packing fractions $\eta=0.556$ and 0.557, respectively, exhibit barrier heights that differ by approximately $1.5k_{B}T$. The barrier height at fixed $\eta$ thus increases with the particle anisotropy. The effect of the particle anisotropy on the effective potential can also be illustrated by comparing the barriers for the systems $S_{2}$ and $S_{6}$ with $L^{*}=3.4$ and $P^*=3.00$ ($\blacktriangle$) and $L^{*}=5.0$ and $P^*=2.00$ (\CIRCLE), respectively. In this case, although the packing fractions are significantly different ($\eta$ = 0.568 and 0.557), both barrier heights are $7.5k_{B}T$. This result shows that a more pronounced particle anisotropy yields significantly higher energy barriers. In conclusion, the barrier height increases with increasing packing fraction and particle anisotropy. Moreover, in our recent work on perfectly aligned hard rods, we noticed that the freezing out of the rotational degrees of freedom has also a quite tangible impact on the height of the barriers, which were found to be higher than those observed in systems of freely rotating rods, especially at low packing fractions \cite{matena}. Such barriers fade out gradually by approaching the continuous N-Sm transition, while they remain finite in the case of the first order N-Sm transition of freely rotating hard rods. One might expect that structural defects, such as screw dislocations and stacking faults, could facilitate the layer-to-layer diffusion by creating barrier-free nematic-like pathways through the layers \cite{selinger}. The small size of our system and the periodic boundary conditions preclude the development of such defects.

The periodic shape of the effective potential determines a hopping-type diffusion along the direction parallel to the nematic director $\hat{n}$, with the particles rattling around in their original layer until they find the appropriate conditions to overcome the barrier and jump to a neighboring smectic layer. An efficacious way to quantify the diffusion of rattling and jumping rods along $\hat{n}$ is provided by the computation of the self part of the VHFs defined in Eq. (2). In Fig. 3, we show the self VHFs for the six systems of interest as a function of \textit{z} at several equidistant times \textit{t}. We detect the appearance of peaks at well-defined locations corresponding to the center-of-mass of the smectic layers along $\hat{n}$, in agreement with previous experimental \cite{lettinga} and theoretical \cite{bier} results. For each aspect ratio, we note that the height of the peaks is larger and the number of peaks is generally higher at the lowest packing fraction. This indicates that decreasing the pressure leads to a higher number of jumping particles which are able to diffuse longer distances. These \textit{fast} particles determine the heterogeneous dynamics of the system and affect its structural relaxation.

\begin{figure}[!ht]
\center
\includegraphics[width=0.48\textwidth]{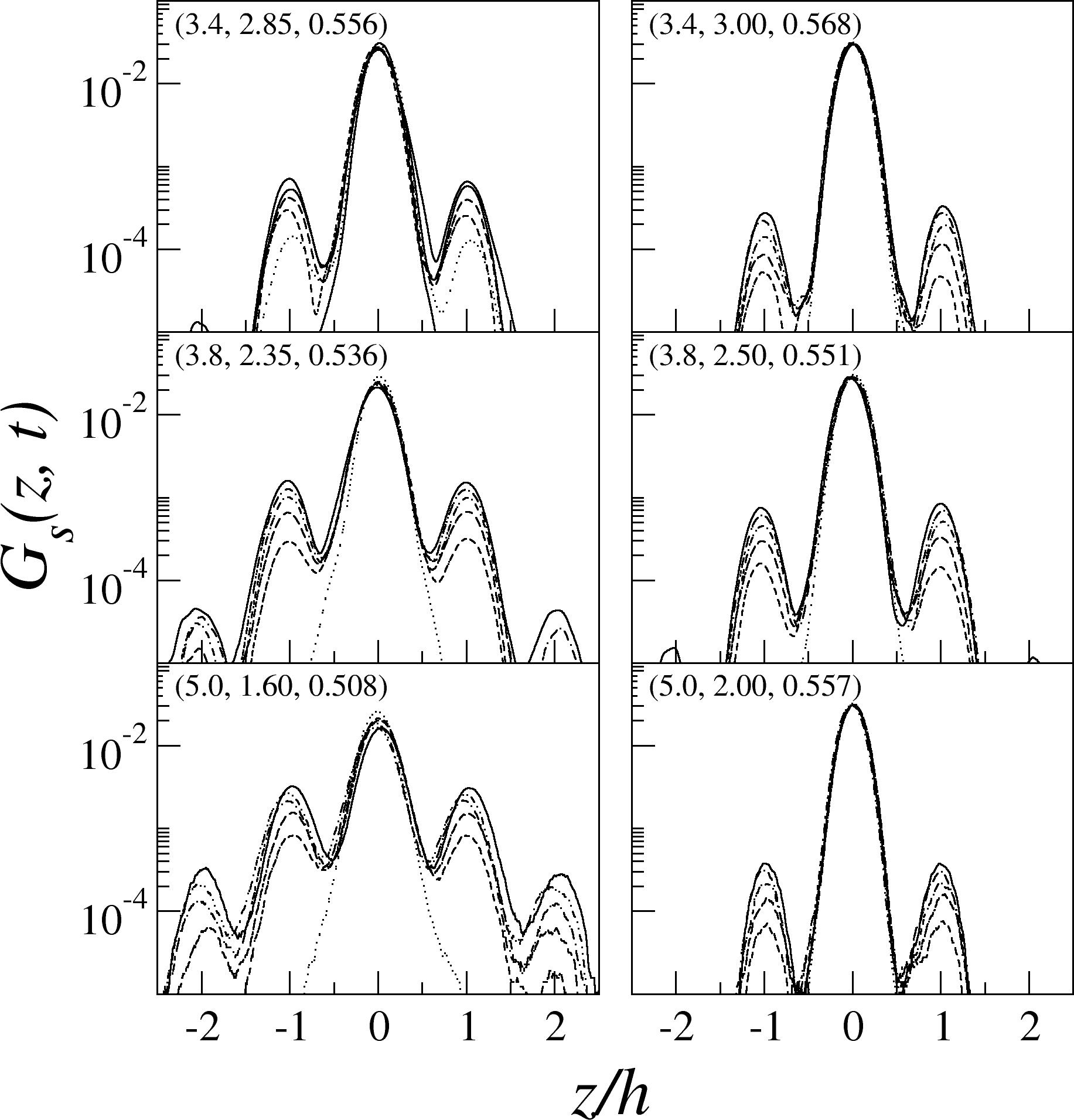}
\caption{Self part of the van Hove function $G_{s}(z, t)$ for smectic LCs of hard spherocylinders with varying length-to-diameter ratios $L^*$ and reduced pressures $P^*$. Length-to-diameter ratio, pressure, and packing fraction are given in each frame as ($L^{*}$, $P^{*}$, $\eta$). The curves refer to the time evolution, from $t=0.4\tau$ (dotted lines) to $t=40\tau$ (solid lines) with increments of $\approx8\tau$.}
\end{figure}

Deviations from the Gaussian behavior of the VHFs have been extensively analyzed in liquid \cite{hurley, kob}, glassy \cite{chaudhuri, vorselaars}, and liquid crystalline \cite{bier, grelet} systems in terms of the non-Gaussian parameter defined in Eq. (3). The NGPs as measured in our six systems are shown in Fig. 4. The in-layer NGPs, $\alpha_{2, xy}$, are essentially negligible for the whole time range, implying a Gaussian in-layer dynamics which is typical of a liquid-like system. The layer-to-layer NGPs exhibit a time-dependent behavior, which is strictly linked to the caging effect exerted on the rods by their nearest neighbors. More specifically, $\alpha_{2, z}$ is basically zero at short times when the rods are rattling around their original location and do not perceive the presence of the surrounding cage. Between $t/\tau=0.1$ and 1.0, $\alpha_{2, z}$ starts to increase, indicating the development of dynamical heterogeneities. During this time interval, the motion of the rods is hampered by the trapping cages and becomes subdiffusive. It is reasonable to assume that the average life-time of the cages corresponds to the time between $\alpha_{2, z}\simeq0$ and $\alpha_{2, z}=\alpha^{max}_{2, z}$, the maximum value of the NGP, which occurs at time $t=t_{max}$ with $2\lesssim t_{max}/\tau \lesssim 10$, depending on the system. This peak increases with  packing fraction and its location determines the beginning of the long-time diffusive regime, where the deviations from  Gaussian behavior start to decrease. Increasing the pressure affects the \textit{z}-diffusion in a twofold manner: (\textit{i}) it increases its heterogeneous behavior, and (\textit{ii}) it delays the onset of the long-time diffusive behavior. Comparison of  the two systems $S_{1}$ and $S_{6}$, corresponding to very similar packing fractions, shows that the dynamics becomes more non-Gaussian for system $S_{6}$ consisting of longer rods and higher energy barriers. On the other hand, for systems  $S_{2}$ and $S_{6}$, displaying similar barrier heights, the non-Gaussian dynamics is again more pronounced for system $S_{6}$ with longer particles. Hence, increasing the anisotropy of the particles yields higher energy barriers and dynamics that is more heterogeneous and non-Gaussian.

\begin{figure}[!ht]
\center
\includegraphics[width=0.48\textwidth]{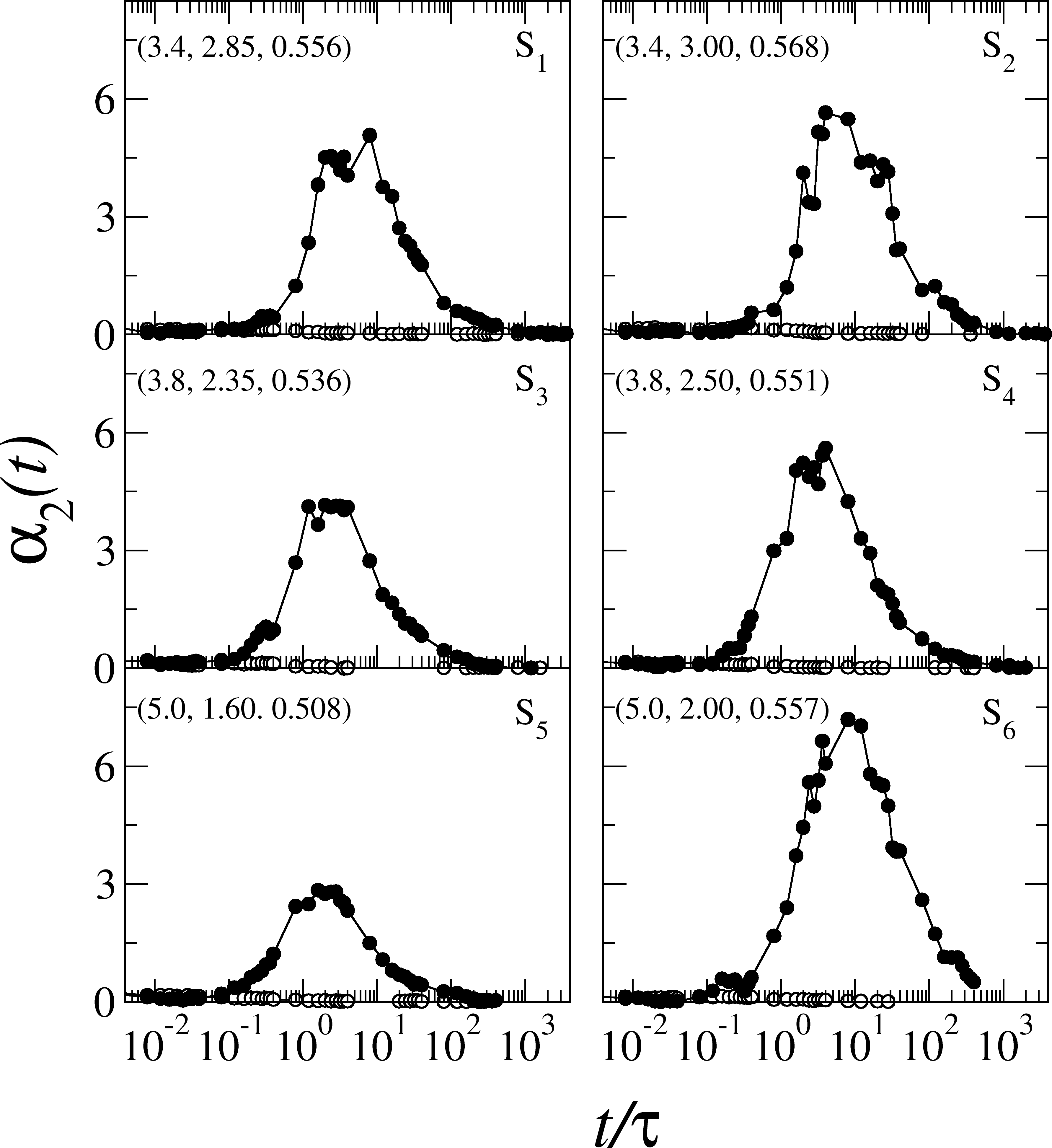}
\caption{Non-Gaussian parameter $\alpha_{2}(t)$ for the layer-to layer (\CIRCLE) and in-layer (\Circle) diffusion for smectic LCs of hard spherocylinders with varying length-to-diameter ratios $L^*$ and reduced pressures $P^*$. Length-to-diameter ratio, pressure, and packing fraction are given in each frame as ($L^{*}$, $P^{*}$, $\eta$). }
\end{figure}

Most of the information obtained by the analysis of the non-Gaussian parameter can also be deducted by the mean square displacements (MSDs), $\left\langle \Delta z^{2}(t) \right\rangle$ and $\left\langle \Delta x^{2}(t)+\Delta y^{2}(t) \right\rangle$, shown in Fig. 5. The \textit{xy}-MSD (open circles) is characterized by a relatively smooth crossover from the short- to long-time diffusion, as observed in slightly dense liquids. By contrast, for the \textit{z}-MSD (solid circles) one clearly detects a more sophisticated behavior as three separate time regimes can be identified. The short-time dynamics, with the rods still rattling in their cages, are diffusive, that is $\left\langle \Delta z^{2}(t) \right\rangle \propto t$. In this regime, $\left\langle \Delta z^{2}(t) \right\rangle > \left\langle \Delta x^{2}(t)+\Delta y^{2}(t) \right\rangle$ because of the anisotropy of the rods \cite{doi}. After an induction time, we observe the formation of a plateau which extends up to $t_{max}$ and quantifies the time to escape from the trapping cages. In this time window, the dynamics becomes subdiffusive. Finally, at different times,  the \textit{xy}-MSD and \textit{z}-MSD become linear with time and the long-time diffusive regime is reached.

\begin{figure}
\center
\includegraphics[width=0.48\textwidth]{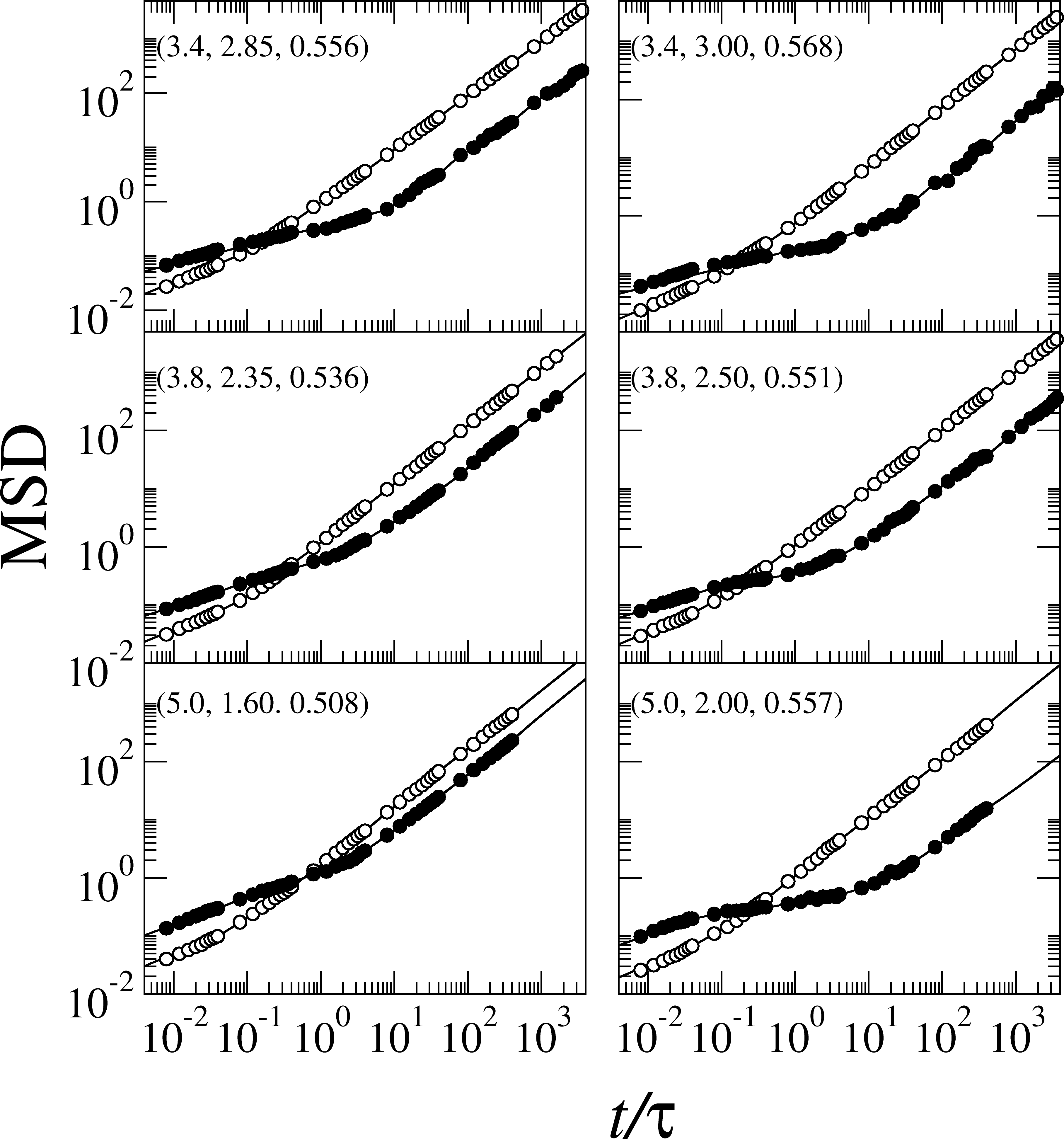}
\caption{Mean square displacement (MSD) in units of $D^{2}$ along the nematic director $\hat{n}$ (solid circles) and along the plane perpendicular to $\hat{n}$ (open circles) for smectic LCs of hard spherocylinders with varying length-to-diameter ratios $L^*$ and reduced pressures $P^*$. Length-to-diameter ratio, pressure, and packing fraction are given in each frame as ($L^{*}$, $P^{*}$, $\eta$). The solid lines are a guide for the eye.}
\end{figure}

From the MSDs in the diffusive regime, we computed the long-time diffusion coefficients in the \textit{z}- and \textit{xy}-directions by applying the well-known Einstein relation \cite{frenkel}. The values of the long-time diffusion coefficients,  $D^L_{xy}$ and  $D^L_{z}$, are presented in Table I in units of $D^{2}/\tau$. The dynamics of each system is characterized by a diffusion coefficient in the \textit{xy}-direction that is larger than the one in the \textit{z}-direction. This result is in agreement with the dynamics in thermotropic smectogenic LCs \cite{cifelli}, but in contrast with recent experiments on the diffusion of \textit{fd} viruses \cite{lettinga}, most probably because of their huge aspect ratio ($L^{*}>100$). In Fig. 6, we exemplarily show the dependence on the packing fraction of the diffusion coefficients and diffusion ratio, defined as 

\begin{equation}
\gamma_{0} \equiv \frac{ D^L_{xy}/D^{2} } { D^L_{z}/(L+D)^{2} } ,
\end{equation}

for  $L^{*}=5.0$. $D^L_{xy}$ and $D^L_{z}$ as well as $\gamma_{0}$ are  well fitted by power law functions of the type $\eta^{-\nu}$, with $\nu \simeq$ 4.6, 26.0, and -21.4, respectively. Increasing the pressure has a significantly larger effect on the inter-layer dynamics than on the in-layer one, as $D^L_{z}$ decreases much faster than $D^L_{xy}$. This is to be expected as the energy barriers (see  Fig. 2) hamper the diffusion more in dense systems. For the lowest packing fractions, i.e., where the smectic phase almost coexists with the nematic phase, the two diffusion coefficients approach each other. Similar considerations are also valid for the systems with aspect ratio $L^{*}=3.4$ and 3.8 (not shown here).

\begin{figure}
\center
\includegraphics[width=0.48\textwidth]{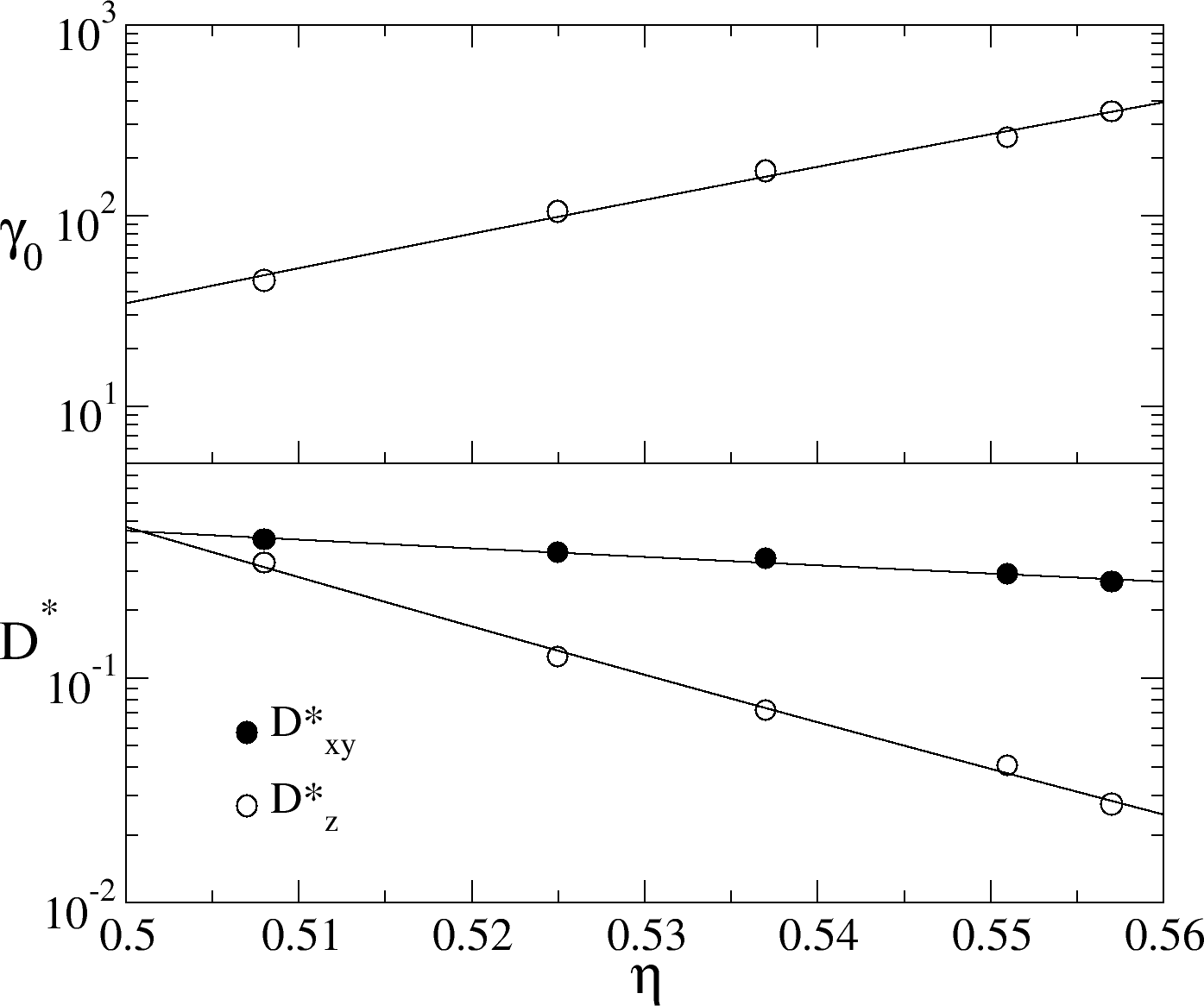}
\caption{Diffusion ratio $\gamma_{0}$ (top) and dimensionless diffusion coefficients (bottom) as a function of the packing fraction for smectic LCs of hard spherocylinders with length-to-diameter ratio  $L^{*}=5.0$. $D^{*}_{xy,z}=D^L_{xy,z}\tau/D^{2}$ denotes the reduced diffusion coefficients, as given in Table I. The solid lines are power law fits.}
\end{figure}

In addition, we analyzed the structural relaxation by calculating the self-part of the intermediate scattering functions, defined in Eq. (4). In Fig. 7, we show $F_{s, xy}(t)$ and $F_{s, z}(t)$ at the wave vectors $\textbf{q} = (q_{x}, q_{y}, 0 )$ and $(0, 0, q_{z})$, respectively, corresponding to the main peaks of the static structure factor. We found that $D\sqrt{(q^{2}_{x}+q^{2}_{y})} \simeq 6$ for all systems, whereas $Dq_{z}\simeq 1.4$, 1.3, and 1.0 for the systems with $L^*= 3.4, 3.8,$ and 5.0, respectively. Regardless the aspect ratio, we can affirm that the in-layer structural relaxation is several orders of magnitude faster than the inter-layer relaxation. If we define the relaxation time $t_{r}$  as the time at which $F_{s}(t)$ decays to $e^{-1}$, then $t_{r, xy}/\tau$ is of the order of $10^{-1}$ and $t_{r, z}/\tau > 10^{3}$. We also find that $F_{s, xy}(t)$ decays very fast to zero with slightly stretched exponential decay, as expected for dense liquid-like dynamics \cite{brambilla}.

By contrast, the inter-layer relaxation develops in two steps separated by a plateau, the beginning of which corresponds to the development of the cage regime. During the initial decay, which is relatively fast ($t/\tau \leq 1$), the rods are free to rattle inside the temporary cage formed by the nearest neighbors \cite{bier}, without perceiving their trapping effect, and $F_{s, z}(t)$ is characterized by an exponential decay. After this short time lapse, we detect a plateau, whose height and temporal extension depend on the packing of the system, as was also found in colloidal glasses \cite {brambilla}. In our previous analysis, we observed that the Gaussian approximation of the self-intermediate scattering function, that is $F^{G}_{s, z}(t)=\exp \left[ -q^{2}_{z}\left\langle \Delta z^{2} (t)\right\rangle \right] $, does not show any significant plateau \cite{patti}. This result indicates that the existence of a plateau must be linked to the non-vanishing  NGP $\alpha_{2, z}$, and hence to the heterogeneous inter-layer dynamics. After the plateau, a second decay, which corresponds to the escape from the temporal cages, leads the systems towards the structural relaxation on a time scale which is long at the highest packing fractions. We were only able to estimate the long-time relaxation decay in the \textit{z}-direction for systems $S_{3}$ and $S_{5}$, due to their relatively low packing fractions. In particular, we fit the long-time decay of $F_{s, z}(t)$ with a stretched exponential function of the form $\exp \left[ -(t/t_{r})^{\beta} \right] $, with $t_{r}/\tau \cong 2500$ and $\beta \cong 0.8$ for $S_{3}$, and $t_{r}/\tau \cong 650$ and $\beta \cong 0.6$ for $S_{5}$. The relaxation time of the remaining systems are beyond our simulation time.

\begin{figure}
\center
\includegraphics[width=0.30\textwidth]{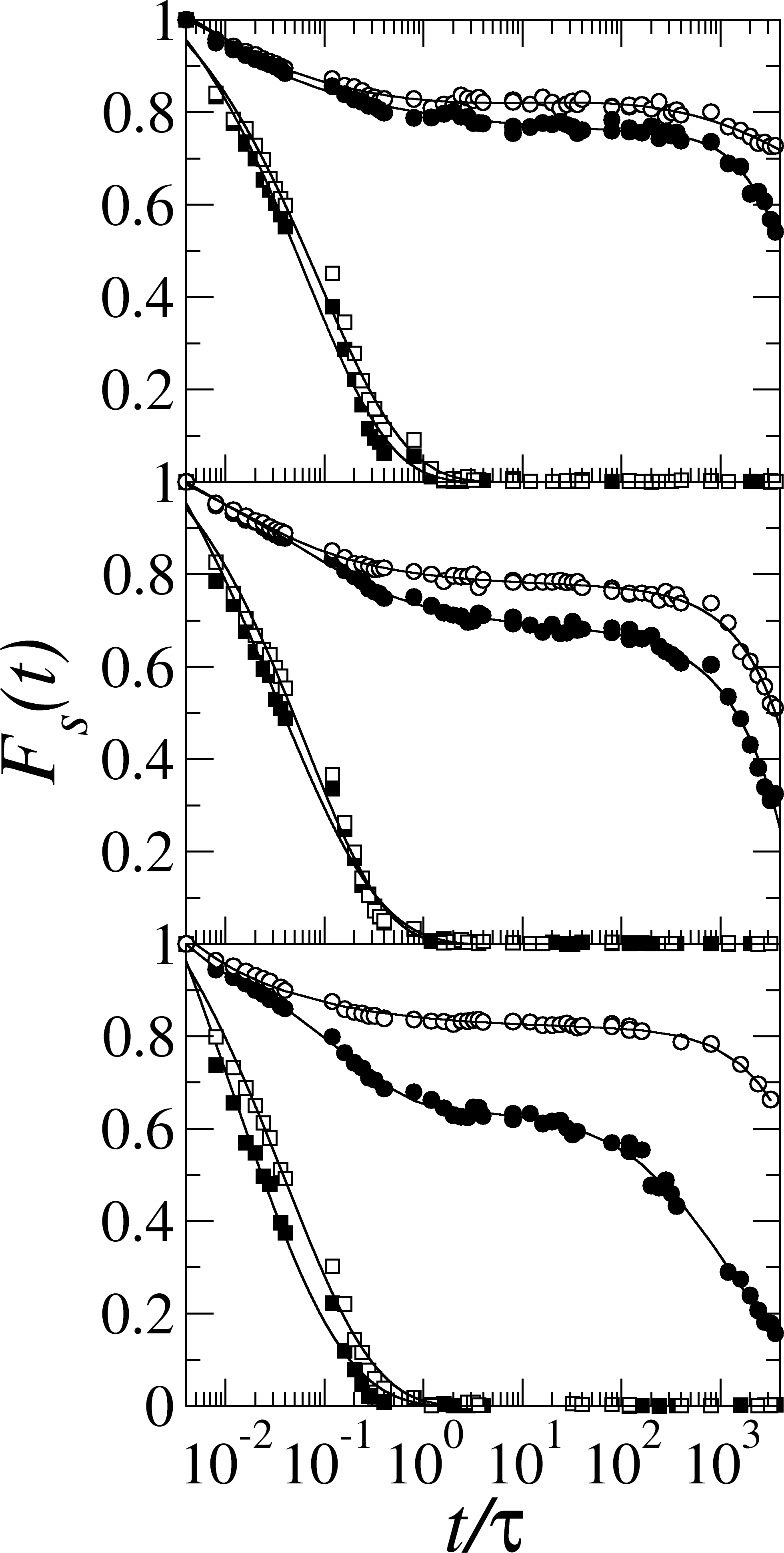}
\caption{Self-intermediate scattering function for smectic LCs of hard spherocylinders with length-to-diameter ratio $L^{*}=3.4$, and reduced pressures $P^*=2.85$ and 3.00 (top), $L^{*}=3.8$, and $P^*=2.35$ and 2.50 (middle), and $L^{*}=5.0$, and $P^*=1.60$ and 2.00 (bottom).  The solid and open symbols refer to the lowest and highest pressure, respectively. Squares and circles refer, respectively, to the in-layer and inter-layer relaxation. The solid lines are fits.}
\end{figure}

The results shown so far give clear evidence of the existence of \textit{fast-moving} particles determining a distribution of decay rates which affects the long-time structural relaxation. We now turn our attention to the occurrence of \textit{collectively} moving particles which might play a crucial role, or might even be  responsible, for the heterogeneous inter-layer dynamics in smectic LCs. To this end, we first label the fast-moving particles by applying the static cluster criterion defined below Eq. (5). Fig. 8 shows the probability size distribution $P(n)$ of the number of interlayer rods  \textit{n} in a string-like cluster using rattling distance $\delta_{rat}=2\sigma$, where $\sigma$ is the standard deviation  as specified in Eq. (5) and presented in Table I. In Fig. 9, we give an illustrative example of static strings observed in system $S_{3}$. Regardless the particle anisotropy, two interesting conclusions can be drawn: (\textit{i}) the observed strings consist mostly of 2 or 3 rods, while clusters of more than 5 rods are extremely rare, but do exist; and (\textit{ii}) strings containing more than 2 rods are more often formed in the denser state, as usually observed in supercooled liquids and glassy systems, where the average cluster size increases when the caging effect becomes stronger \cite{donati}. The semilog plot of Fig. 8 shows that the size probability distribution is roughly exponential. In particular, we observed that $P(n) \propto \exp \left(  -\alpha n \right) $, from which the average cluster size can be estimated: $\left\langle n \right\rangle = \left( 1-\exp(-\alpha)\right) ^{-1}$. The largest clusters are thus found for the shortest rods and the highest pressure.

\begin{figure}
\center
\includegraphics[width=0.30\textwidth]{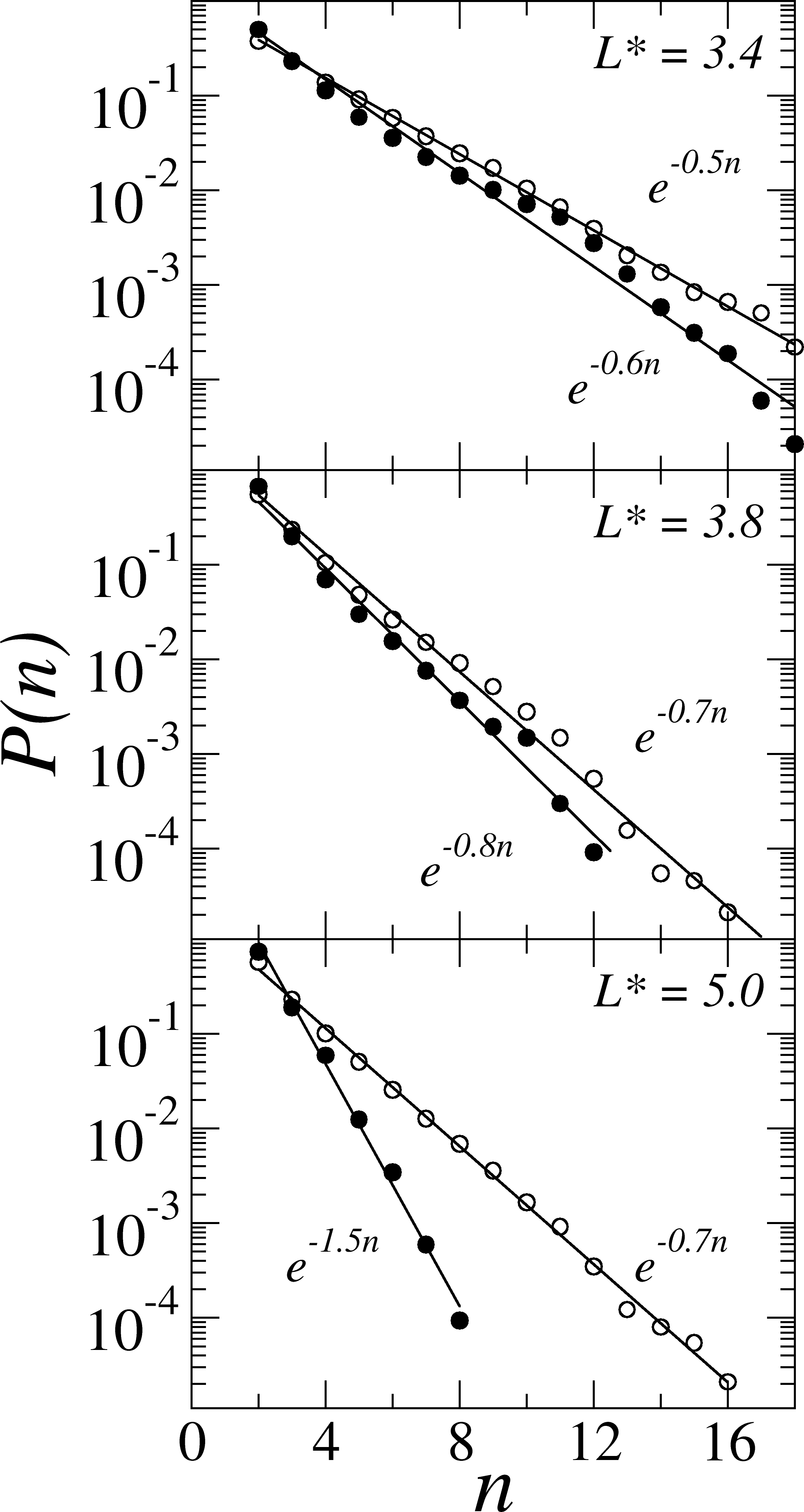}
\caption{Size probability distribution $P(n)$ of the number of interlayer rods  $n$ in a stringlike clusters (with $\delta_{rat}=2\sigma$, see text), in  a smectic LCs of hard spherocylinders with length-to-diameter ratio  $L^{*}=3.4$, and reduced pressures $P^*=2.85$ and 3.00 (top), $L^{*}=3.8$, and $P^*=2.35$ and 2.50 (middle), and $L^{*}=5.0$, and $P^*=1.60$ and 2.00 (bottom). The solid and open symbols refer to the lowest and highest pressure, respectively. The solid lines denote the fit $P(n) \propto \exp (-\alpha n)$, with $\alpha$ given in the figure.}
\end{figure}

\begin{figure}
\center
\includegraphics[width=0.20\textwidth]{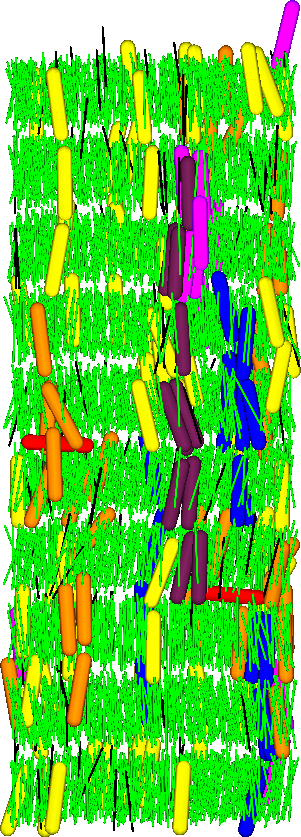}
\caption{(color online). Snapshot of 3000 rods with length-to-diameter ratio $L^{*}=3.8$ and packing fraction $\eta=0.536$. In-layer rods and single interlayer rods (black), both with diameters reduced to $D/4$ for clarity, are predominantly shown. The thicker rods denote transverse ones (red or dark shaded) as well as stringlike clusters of 2 (yellow or light shaded) up to 11 (brown or dark shaded) rods.}
\end{figure}

The energy barriers of Fig. 2 and the periodically peaked shape of the VHFs of Fig. 3 unfold the rattling-and-jumping layer-to-layer diffusion of the rods, but only provide a global picture of the dynamics in smectic LCs. In order to gain a deeper understanding of the actual dynamics on the particle scale behind the layer-to-layer hopping-type diffusion, we followed the trajectories of single particles. Interestingly, we observed that some rods diffuse very fast, others move to the inter-layer spacing, where they might reside for a long time, and then return to their original layer, and others move from one layer to another several times or perform consecutive jumps as shown in Fig. 10b for system $S_{3}$. Furthermore, transverse inter-layer rods, although extremely rare due to the high energy barriers \cite{vanduijneveldt, vanroij}, have also been  detected. These transverse inter-layer particles might diffuse either to a new layer or go back to the old one by keeping or changing their original orientation. This variegated behavior suggests a rather broad distribution of layer-to-layer jump times $\Pi(t_{J})$, where $t_{J}$ is the time between the first and last contact with the new and old layer, respectively. Such a contact is established as soon as the particle is at a rattling distance $\delta_{rat}=2\sigma$ from the middle of a smectic layer. The probability distributions of jump times for systems $S_{1}-S_{6}$ have been computed by averaging over at least \textit{N} jumps, with \textit{N} the number of rods in each system. In Fig. 10a, we give  the  probability distribution of jump times $\Pi(t_{J})$ for  system $S_{3}$. As expected, the distribution is not particularly narrow, but extends over two time decades $0.01< t_{J}/\tau <1$, with the most probable jump time \textit{\={t}}=0.17$\tau$ and the median time (i.e. the time at which 50\% of the jumps have been performed) $t^{*}_{J}=0.28\tau$. These times increase at $\eta=0.551$, as given in Table I. The remaining systems with shorter or longer rods show the same tendency and similar times (see Table I). We also distinguished  single and consecutive (or multiple) jumps according to the dwelling time being longer or shorter, respectively, than $t^{*}_{J}$ before  another jump is started. Single jumps are significantly more frequent than  multiple jumps, especially at high densities where the latter are less than 1\% of the total number of jumps. Most of the multiple jumps consist of double or triple jumps, while quadruple ones are basically irrelevant. The largest number of multiple jumps was observed in the system with the lowest packing fraction, i.e., $S_{5}: L^*=5.0, \eta=0.508$, where $\approx5\%$ of the total jumps is multiple, of which $\approx83\%$ is a double and the remaining fraction is a triple jump.

\begin{figure}
\center
\includegraphics[width=0.44\textwidth]{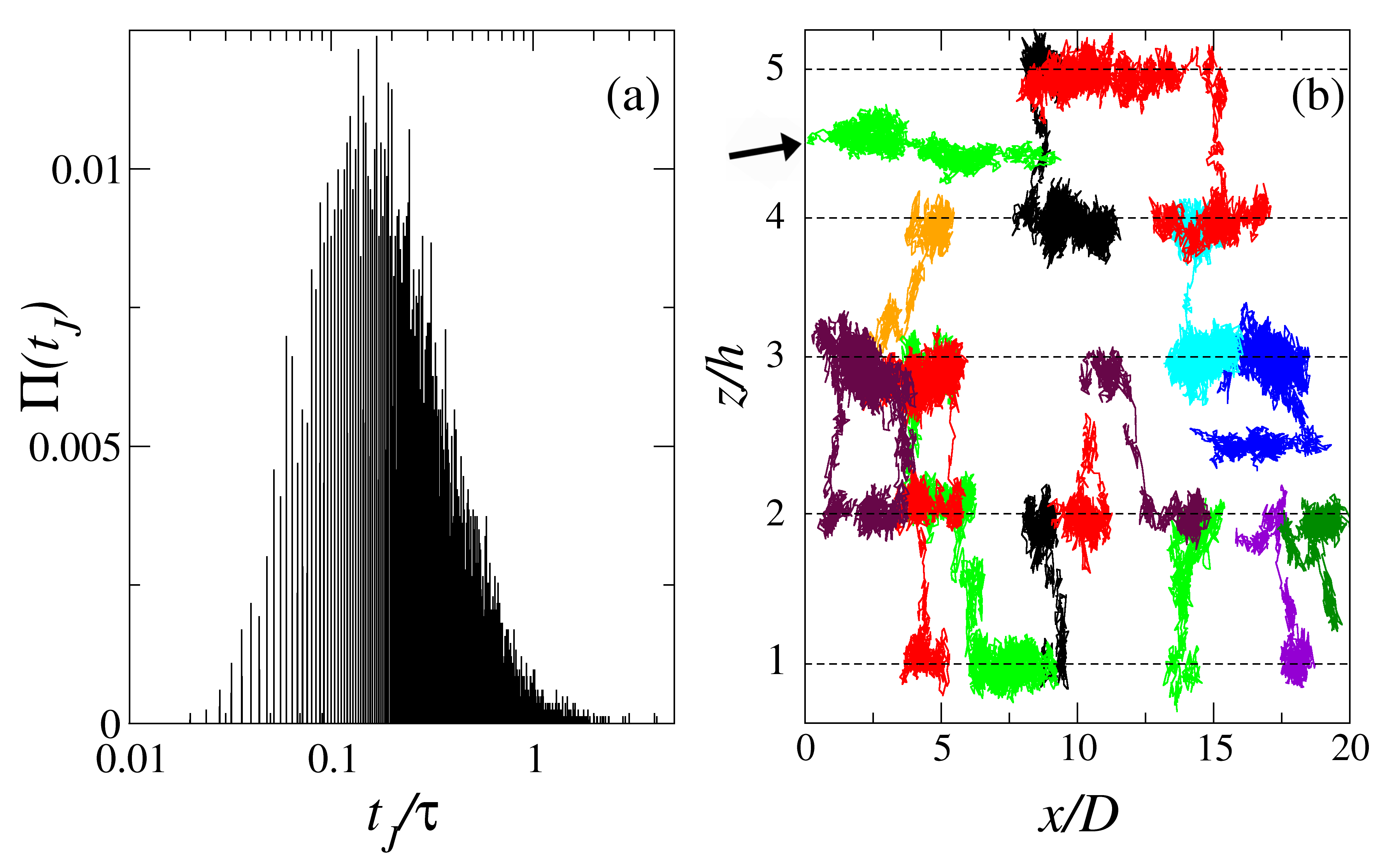}
\caption{(color online). (a) Probability distribution of jump times $\Pi(t_{J})$ based on $\delta_{rat}=2\sigma$ for lenght-to-diameter ratio $L^{*}=3.8$ and packing fraction $\eta=0.536$. (b) Trajectories of jumping rods in the same system projected onto the \textit{xz} plane, with the dashed lines locating the  centre of the smectic layers. The arrow indicates the trajectory of a transverse interlayer particle.}
\end{figure}

The formation of \textit{static} stringlike clusters does not necessarily imply the occurrence of dynamic cooperativity, as a rod that belongs to a cluster might still   diffuse  individually from layer-to-layer or might fail to jump. In order to ascertain the occurrence of collective motion of strings, we introduced a \textit{dynamic} cluster criterion. More specifically, we assume that two jumping rods \textit{i} and \textit{j} are actually moving cooperatively, if (1) their arrival times $t^{(i)}$ and $t^{(j)}$ in their new layers satisfy the condition $\vert t^{i}-t^{j}\vert < \Delta t_{0}$ and (2)  their \textit{z} and \textit{xy} distances at $t^{(i)}$ and $t^{(j)}$ are smaller than \textit{h} and \textit{D}, respectively. To select a consistent value for $\Delta t_{0}$, we use the distribution of jump times $\Pi(t_{J})$. If we assume that $\Delta t_{0}$ is the maximal jump time $t^{max}_{J}$ (i.e. long enough for all jumps to be performed according to $\Pi(t_{J})$), we find that for $L^{*}=3.4$ the ratio between the number of collective jumps and the total number of jumps is $f_{c}^{max}$ as given in Table I. These jumps involve mainly two or three rods, whereas collective jumps of 4 or more rods are extremely rare. We further note that the vacated space of a jumping rod can be either occupied by another rod jumping in the same direction or by a rod jumping in the opposite direction, with roughly the same probability. A less restrictive spatial criterion would not affect significantly these values. By contrast, the number of collective jumps is rather sensitive to the temporal criterion. If we reduce $\Delta t_{0}$ to the median jump time $t^{*}_{J}$, then the fraction of collective jumps, $f^{*}_{c}$, decreases substantially as shown in Table I. Regardless the details of the cluster criterion, we thus find a fraction of $10^{-2}-10^{-1}$ of the jumps to be collective, the more so for longer rods at lower pressures.

The motion is indeed strongly cooperative at  low packing fractions, despite the larger static stringlike clusters detected in the denser systems (see Fig. 8). This result is most probably due to the permanent smectic barriers which increase upon approaching the smectic-to-solid phase transition and hence hamper the attempted jumps of the rods in the strings. By contrast, in glass-forming systems, where no permanent barriers are observed, the cluster size increases upon approaching the glass transition \cite{donati, berthier2}.

\section{CONCLUSIONS}
\label{sec:CONCLUSIONS}

In summary,  we have studied the diffusion and structural relaxation  in \textit{equilibrium} smectic LC phases of hard rods with different anisotropies by computer simulations. Remarkably, these systems exhibit non-Gaussian layer-to-layer diffusion and dynamical heterogeneities which are similar to those observed in \textit{out-of-equilibrium} supercooled liquids. The simultaneous presence of temporary cages due to the trapping action of neighboring rods and the permanent barriers due to the static smectic background, provokes a rattling-and-jumping diffusion which influences the long-time structural relaxation decay. In analogy with glassy systems, one can clearly distinguish three separate time regimes for \textit{z}-motion. The short-time diffusive regime, with the rods rattling around their original location without feeling the  surrounding neighbors, is characterized by a Gaussian distribution of the \textit{z}-displacements and an exponential temporal relaxation. The subdiffusive regime at intermediate times displays a non-Gaussianity and a plateau in both mean square displacement and intermediate scattering function. At this stage, the interlayer dynamics is heterogeneous with fast-moving particles diffusing individually or cooperatively in a stringlike fashion. Finally, at long times, the systems enter a second diffusive regime with Gaussian distributions of the displacements and non-exponential decay of the intermediate scattering function. By contrast, at all time-scales, the in-layer diffusion is typical for  two-dimensional fluids with a negligible NGP and a structural relaxation which is at least 4 time decades faster than the inter-layer one.

The analysis of the self-VHFs points out the tendency for the rods to diffuse from layer to layer through quasi-discretized jumps. This hopping-type motion is significantly hampered in very dense systems, where the barriers for the layer-to-layer diffusion intensify the confinement of the rods to the middle of the smectic layers. The temporal extension of the jumps is not uniform, but characterized by a rather broad time distribution which covers approximately two time decades. Depending on the dwelling time between two successive jumps of the same particle, single and multiple jumps have been detected, with the former significantly more frequent than the latter, especially at high densities. Although the inter-layer rods are usually oriented along the nematic director, some of them assume a transverse orientation. The long tails of the VHFs indicate the presence of particles that are able to diffuse much longer distances than the average, especially at low packing fractions. Interestingly, the dynamic behavior of such fast-moving particles supports the intriguing analogy with glassy systems even further. In particular, fast-moving particles assemble in stringlike clusters whose average length increases upon approaching the smectic-to-crystal phase transition. Likewise, fragile and strong glass-formers show a similar tendency when cooled down towards the glass transition temperature \cite{donati, teboul}. We also gave clear evidence that the strings detected in static configurations can promote collective diffusion of jumping particles. This is especially tangible at low packing fractions where the hampering action of the permanent energy barriers is less effective. Finally, we also investigated the effect of particle anisotropy on the non-Gaussian layer-to-layer diffusion and cooperative motion in smectic LCs. We find that at fixed packing fraction, the barrier height increases with increasing particle anisotropy, and hence the dynamics is more heterogeneous and non-Gaussian for longer rods, yielding a lower diffusion coefficient along the nematic director and smaller clusters of inter-layer particles that move less cooperatively. At fixed barrier height,  the dynamics becomes more non-Gaussian and heterogeneous for longer rods; smaller clusters move more collectively, giving rise to  a higher  diffusion coefficient along the nematic director.

Our results, which might be relevant for the study of the dynamics in confined fluids \cite{nugent} as well as for the diffusion of lipids and proteins in cellular membranes \cite{falck, gurtovenko}, are already stimulating the analysis of dynamical processes in columnar liquid crystals, where the presence of inter-columnar energy barriers provokes a non-Gaussian hopping-type motion and a relaxation behavior that is also remarkably similar to that of out-of-equilibrium supercooled liquids \cite{belli}. We hope that our findings stimulate further theoretical and experimental studies of particle-scale dynamics in
heterogeneous (confined, smectic, columnar) liquids.

\begin{acknowledgments}
This work was financed by an NWO-VICI grant. We thank M. Bier and P. van der Schoot for useful discussions.
\end{acknowledgments}

\end{document}